\documentclass[11pt,showpacs,showkeys,preprint,prstab]{revtex4}
\input epsf

\begin{document}

\date{July 5th, 2002}

\title{Sub-picosecond compression by velocity bunching in a photo-injector}

\author{P. Piot,} 
\thanks{Now at Fermi National Accelerator Laboratory}
\email{piot@fnal.gov}
\address{Deutsches Elektronen-Synchrotron (DESY), 
D-22607 Hamburg, Germany} 
\author{L. Carr, W.S. Graves}
\thanks{Now at Massachusetts Institute of Technology}
\email{wsgraves@mit.edu}
\author{H. Loos, }
\address{Brookhaven National Laboratory, Upton, 
NY 11973, USA}

\begin{abstract}
We present an experimental evidence of a bunch compression scheme that 
uses a traveling wave accelerating structure as a compressor. The bunch 
length issued from a laser-driven radio-frequency electron source was 
compressed by a factor $>$3 using an S-band traveling wave structure located 
immediately downstream from the electron source. Experimental data are found 
to be in good agreement with particle tracking simulations.
\end{abstract}

\maketitle

\section{Introduction}
In the recent years there has been an increasing demand on ultrashort electron bunches 
to drive short-wavelength free-electron lasers and study novel accelerating techniques 
such as plasma-based accelerators~\cite{yurkovPRL,barov}. Short bunches are commonly obtained 
by magnetic compression. In this latter scheme, the bunch is compressed using a series 
of dipoles arranged in a chicane configuration such to introduce an energy-dependent pathlength. 
Therefore an electron bunch having the proper time-energy correlation can be shortened in the 
chicane. However, problems inherent to magnetic compression such as momentum spread and 
transverse emittance dilution due to the bunch self-interaction via coherent synchrotron 
radiation~\cite{derbenev} has brought back the idea of bunching the beam with 
radio-frequency (rf) structures~\cite{haimson}. 

It was recently proposed to incorporate the latter method (henceforth named velocity 
bunching) into the next photo-injector designs~\cite{serafini}. The velocity bunching 
relies on the phase slippage between the electrons and the rf-wave that occurs during the 
acceleration of non ultra-relativistic electrons. In this paper after presenting a brief 
analysis of the velocity bunching scheme, we report on its exploration at the deep ultraviolet 
free-electron laser (DUV-FEL) facility of Brookhaven National Laboratory (BNL). The measurements 
are compared with numerical simulations performed with the computer program ASTRA~\cite{astra}.  

\section{Analysis of the velocity bunching technique}
\begin{figure}[h]
\begin{center}
\epsfxsize=160mm
\epsfbox{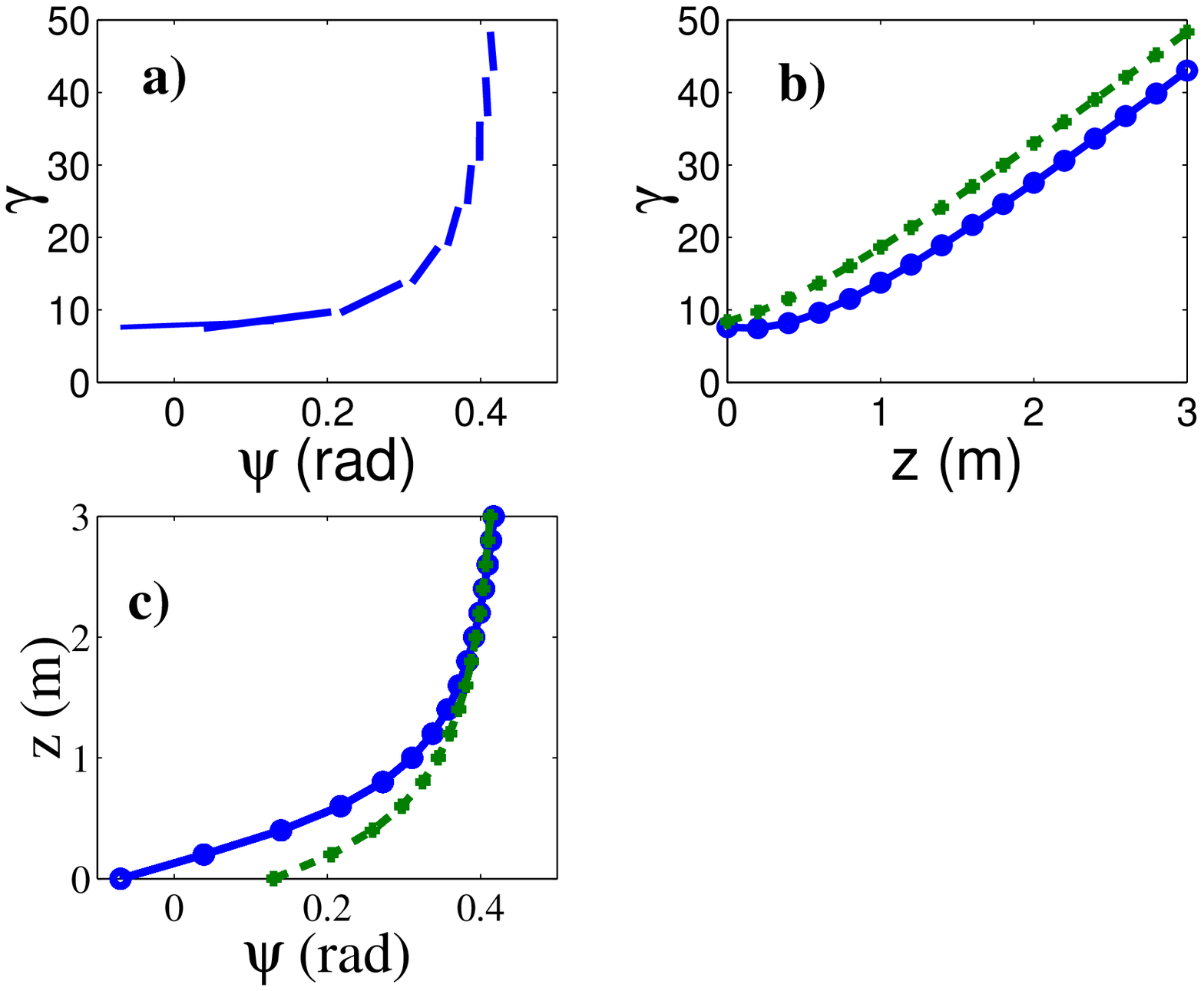}
\end{center}
\caption{ Simulation, using a two macro-particles model, of the velocity compression 
in a 3~m long traveling wave structure. The initial conditions are $(\psi_o, \gamma_o)$=(0,8) 
and the macro-particle spacing is $(\Delta \psi_o, \Delta \gamma_o)$=(0.1, 0.4). 
Plot {\bf a)} shows snapshots at different $z$ of the longitudinal phase space each segment 
extremities is determined by the two macro-particles positions. 
Plots {\bf b)} and {\bf c)} present the energy gain and phase evolution of the two 
macro-particles versus $z$. In these two latter plots, solid lines represent the leading 
particle and dashed lines the trailing one.}
\label{fig:theory}
\end{figure}

In this section we elaborate a simple model that describes how the velocity 
bunching works. A more detailed discussion is given in Reference~\cite{serafini}.\\
An electron in an rf traveling wave accelerating structure experiences the longitudinal 
electric field:
\begin{eqnarray}
E_z(z,t)=E_o \sin(\omega t -k z + \psi_o) \mbox{,}
\end{eqnarray}
where \( E_{o} \) is the peak field, \( k \) the rf wavenumber and \( \psi_{o} \) the
injection phase of the electron with respect to the rf wave. 
Let \( \psi (z,t)=\omega t-kz+\psi _{o} \) be the relative phase of the electron w.r.t the 
wave. The evolution of \( \psi (t,z) \) can be expressed as a function of \( z \) solely:
\begin{eqnarray} \label{eqn:psi}
\frac{d\psi }{dz}=\omega \frac{dt}{dz}-k=\frac{\omega }{\beta c}-k=
k\left( \frac{\gamma }{\sqrt{\gamma ^{2}-1}}-1\right) \mbox{.}
\end{eqnarray}
Introducing  the parameter \( \alpha \doteq \frac{eE_{o}}{k mc^{2}} \), we write for the 
energy gradient~\cite{kim}: 
\begin{eqnarray} \label{eqn:gammap}
\frac{d\gamma }{dz}= \alpha k \sin (\psi ) \mbox{.}
\end{eqnarray}
The system of coupled differential equations (\ref{eqn:psi}) and (\ref{eqn:gammap}) 
with the initial conditions $\gamma(z=0)=\gamma_o$ and $\psi(z=0)=\psi_o$ describe 
the longitudinal motion of an electron in the rf structure. Such a system is  solved 
using the variable separation technique to yield:
\begin{eqnarray} \label{eqn:res2}
\alpha \cos \psi + \gamma -\sqrt{\gamma^2-1} = {\cal C}\mbox{.}
\end{eqnarray}
Or, expliciting $\psi$ as a function of $\gamma$:
\begin{eqnarray} \label{eqn:res3}
\psi(\gamma)=\arccos \left(\frac{{\cal C}-\gamma +\sqrt{\gamma^2-1}}{\alpha}
\right)
\mbox{.}
\end{eqnarray}
Here the constant of integration is set by the initial conditions of the 
problem\footnote{The constant ${\cal C}$ defined in Eq.~\ref{eqn:res2} corresponds  
to the Hamiltonian defined in Ref.~\cite{serafini} evaluated for a wave with 
velocity $v=c$, where c is the velocity of light.}: 
${\cal C}=\alpha \cos \psi_o +\gamma_o -\sqrt{\gamma_o^2-1}$. The latter equation 
gives insights on the underlying mechanism that provides compression. In order to 
get a simpler model, we consider the limit:  
$\psi_{\infty}\doteq \lim_{\gamma \rightarrow \infty} \psi(\gamma) = 
\mbox{arccos}\left(\cos (\psi_o) +\frac{1}{2\alpha \gamma_o} \right)$; we have assumed 
$\gamma_o$ is larger than unit and did the approximation 
$\gamma_o -\sqrt{\gamma_o^2-1}\simeq 1/(2\gamma_o)$. After differentiation 
of Eq.~\ref{eqn:res3}, given an initial phase $d\psi_o$ and energy $d\gamma_o$ extents  
we have for the final phase extent:
\begin{eqnarray} \label{eqn:res4}
d\psi_{\infty}=\frac{\sin (\psi_o)}{\sin (\psi_{\infty})} d\psi_o + 
\frac{1}{2\alpha \gamma_o^2 
\sin(\psi_{\infty})} d\gamma_o 
\mbox{.} 
\end{eqnarray}
Hence depending upon the incoming energy and phase extents, the phase of injection 
in the rf structure $\psi_o$ can be tuned to minimize the phase extent after 
extraction i.e. to ideally (under single-particle dynamics) make $d\psi_{\infty} \rightarrow 0$. 
We note that there are two contributions to $d\psi_{\infty}$: the first term 
$\partial \psi_{\infty} / \partial \psi_o$ comes from the phase slippage (the injection 
and extraction phases are generally different). The second term 
$\partial \psi_{\infty} / \partial \gamma_o$ is the contribution coming from the initial 
energy spread. To illustrate the compression mechanism we consider a two macro-particles 
model. In Figure~\ref{fig:theory} we present results obtained by numerically integrating the 
equation of motion for two non-interacting macro-particles injected into a 3~m long 
traveling wave structure. Given the incoming phase $\Delta \psi_o$ and energy $\Delta \gamma_o$ 
spreads between the two macro-particles, and the accelerating gradient of the structure 
(taken to be 20~MV/m), we can optimize the injection phase to minimize the bunch length at the 
structure exit.  

\section{Experimental results}
\begin{figure}[h]
\begin{center}
\epsfxsize=160mm
\epsfbox{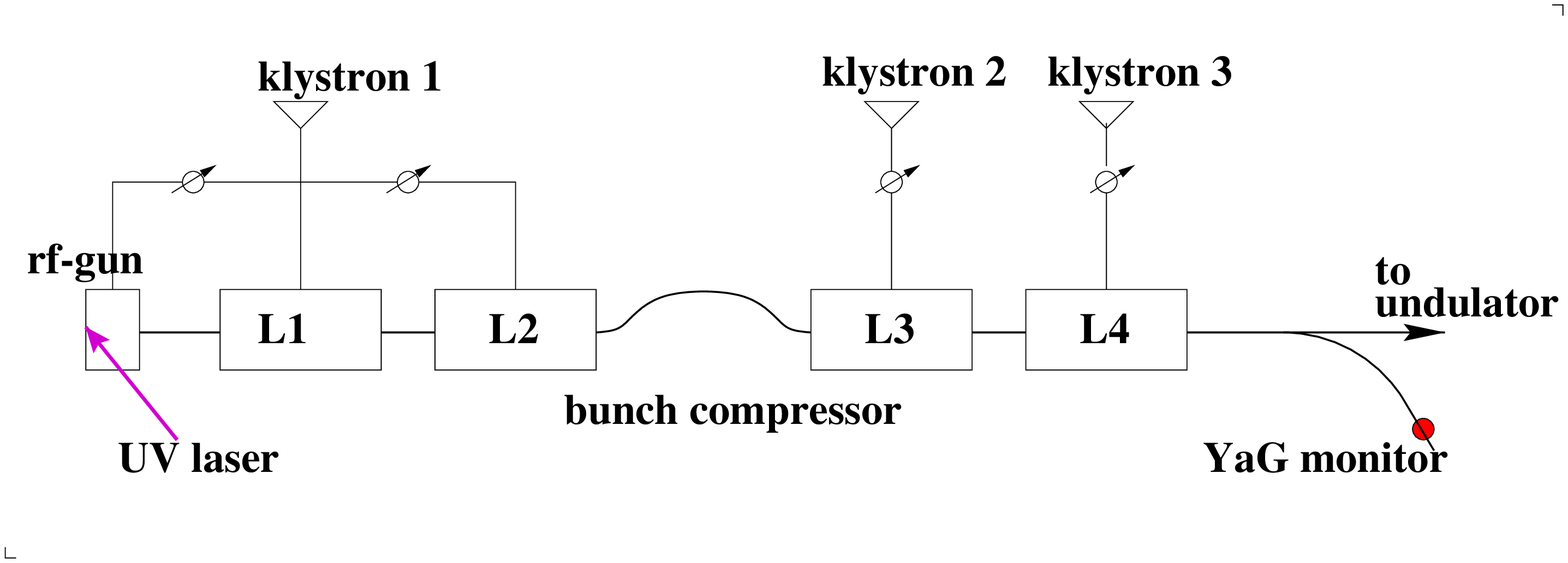}
\end{center}
\caption{ Overview of the Deep ultra-violet free-electron laser (DUV-FEL) accelerator. 
L1, L2, L3, and L4 are the four linac sections.}
\label{fig:duvfel_pic}
\end{figure}

The measurement was carried out at the DUV-FEL facility 
of Brookhaven national laboratory~\cite{duvfel}. A block diagram of the linear accelerator 
is given in Fig.~\ref{fig:duvfel_pic}. The electron bunches of $\sim$4~MeV, generated 
by a laser-driven rf electron source, are accelerated by a series of four linac sections. 
The linac sections consist of 2.856~GHz traveling wave structures operating on the $2\pi/3$ 
accelerating mode. The structures are approximately 3~m long and can operate with an average 
accelerating voltage up to 20~MV/m. Nominally the bunch is shortened using a magnetic 
bunch compressor chicane located between the second and third linac sections. In this 
latter case, the linac sections L1, L3, L4 are ran on-crest while the linac L2 
is operated to impart the proper time-energy correlation along the bunch to enable  
compression as the beam pass through the magnetic chicane. \\

To investigate the velocity bunching scheme, the linac section L1 was used as a 
buncher: its phase was varied and, for each phase setting, the section L2 was 
properly phased to maximize the beam energy with sections L3 and L4 turned off. 
The magnetic bunch compressor was turned off during the measurement. The 
nominal settings for the different rf and photo-cathode drive-laser parameters 
are gathered in Table~\ref{tab:setup}. 
\begin{table}[h!]
\begin{center}
\begin{tabular}{l c c}
\hline \hline 
parameter                       &      value       & units  \\ \hline 
laser injection phase           &  40   $\pm$ 5	   & rf-deg \\
laser radius on cathode         &  0.75	$\pm$ 0.1  & mm     \\
laser rms length                &  1.15	$\pm$ 0.1  & ps     \\
E-peak on cathode               &  83   $\pm$ 1    & MV/m   \\
L1 average accelerating field 	&  10.5 $\pm$ 0.1  & MV/m   \\
L2 average accelerating field 	&  13.2 $\pm$ 0.1  & MV/m   \\
\hline 
\hline 
\end{tabular}
\caption{\label{tab:setup} Nominal settings for the rf-gun, accelerating sections, and the 
photo-cathode drive-laser. The values have been directly measured or inferred from the beam 
properties. }
\end{center}
\end{table} 

The measurements of bunch length that follow are compared with simulations performed 
with the program ASTRA~\cite{astra}. ASTRA is a macro-particle tracking code based on a 
rotational symmetric space charge algorithm. It  incorporates a detailed 
model for the traveling wave accelerating structure~\cite{loew,massimo}. To perform the 
simulations we used the parameters values of Table~\ref{tab:setup}. 
The laser transverse distribution was modeled by a radially uniform transverse 
distribution with 0.75~mm radius, and the time profile, measured using a single shot 
cross-correlation technique, was directly loaded into the simulations.\\

Both time- and frequency-domain techniques were used to characterize the
bunching process as the phase of the linac L1 was varied.

The time-domain charge density  was directly measured using the so-called 
zero-phasing method~\cite{wang,graveszp}. In the present case, we use the 
linac section L3 to cancel the incoming time-energy correlation, and 
operate the linac L4 at zero-crossing
to introduce a linear time-dependent 
energy chirp along the bunch (we have investigated both zero-crossing points). The 
bunch is then directed to a beam viewer (``YaG monitor" in Fig.~\ref{fig:duvfel_pic}) 
downstream from a $72^{\circ}$ angle spectrometer. The viewer, located at a 
dispersion (horizontal) of $\eta=907$~mm, allows the measurement of the bunch energy 
distribution. Unlike in Reference~\cite{wang}, the longitudinal phase space of beams 
issued from an rf electron source is not perfectly linear: because of the longitudinal 
space charge forces, the phase space generally has a third order distortion~\cite{dowell}. 
To analyze the impact of such a distortion on our bunch length measurement method, it is 
interesting to consider the Gaussian normalized longitudinal phase space 
$(s, \delta)$ density:
\begin{eqnarray} \label{eqn:phisp}
{\cal P}(s, \delta)= \frac{1}{2\pi \sigma_{\delta} \sigma_s} \times \exp\left(
-\frac{(\delta-h_1s -h_3 s^3)^2}{2\sigma_{\delta}^2}\right) \times 
\exp\left(-\frac{s^2}{2\sigma_s^2}\right)\mbox{.}
\end{eqnarray}
Here $\sigma_s$ and $\sigma_{\delta}$ are the bunch rms length and rms uncorrelated 
fractional momentum spread and $h_1$, $h_3$ are constants that quantify the linear and third 
order correlations of the longitudinal phase space. The zero-phasing measurement can then be 
analyzed 
in term of a sequence of numerical calculation based on Eq.~\ref{eqn:phisp}: by computing and 
comparing the time and fractional momentum spread projections associated to 
${\cal P}(s, \delta+C_o\times s)$. The constant $C_o$ depends on the incoming beam 
energy $E_o$, the accelerating voltage of the zero-phased linac section, the 
rf wavenumber $k_{rf}$, and dispersion~\cite{wang}: 
$C_o=\pm \frac{E_o}{\eta V_{rf}k_{rf}}$ the $\pm$ sign reflects the two possible 
zero-crossing points. \\

\begin{figure}[h]
\begin{center}
\epsfxsize=160mm
\epsfbox{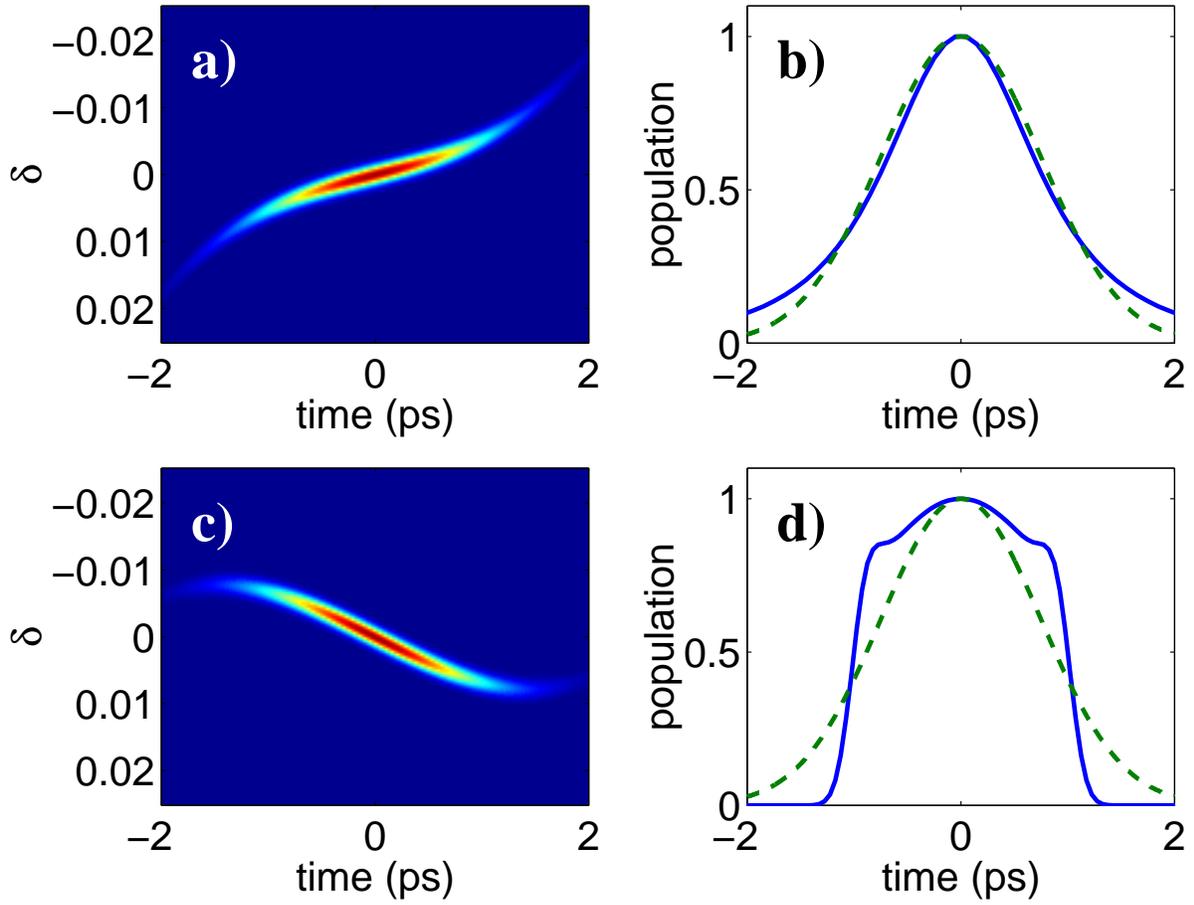}
\end{center}
\caption{ Simulation of the zero-phasing method for a distorted incoming longitudinal phase 
space. The images {\bf a)} and {\bf c)} depict the phase spaces after the bunch as passed the 
zero-phasing traveling wave structure for 
the "positive" (upper plots) and "negative" (lower plots) zero-crossing points. The plots {\bf b)} and {\bf d)} are the 
corresponding projections. In these plots we compare the time projection (dashed lines) with the one 
deduced from the fractional momentum spread projection (solid lines). The time $>0$ corresponds to 
the bunch tail.}
\label{fig:zerophasingsim}
\end{figure}

An example of such a calculation is presented in Fig.~\ref{fig:zerophasingsim}. To generate 
the presented data we started with a longitudinal phase space which has a third order 
distortion but no linear correlation (as it should be downstream from linac L3) we then 
set the constant $C_o$ to have a full-width fractional momentum spread of approximately $1.5$~\% 
similar to the value imposed by the finite size of the viewer (diameter $\sim$15~mm) used for the 
measurement of the bunch energy distribution.  
The Figure~\ref{fig:zerophasingsim} demonstrates the impact of the third order distortion 
in the longitudinal phase space: depending on the chosen zero-crossing phase, it contributes 
to an elongation or a contraction of the measured time profile compared to the real profile. 
For the rms bunch length measurements reported hereafter we computed the average bunch 
length measured for the two zero-crossing points and reported the difference as an error bar. 
For the reported bunch profiles we use the bunch profile corresponding to the case when 
the phase space has no fold over. Hence we expect the bunch time-profile reported hereafter to
be longer than in reality.\\

\begin{figure}[h]
\begin{center}
\epsfxsize=150mm
\epsfbox{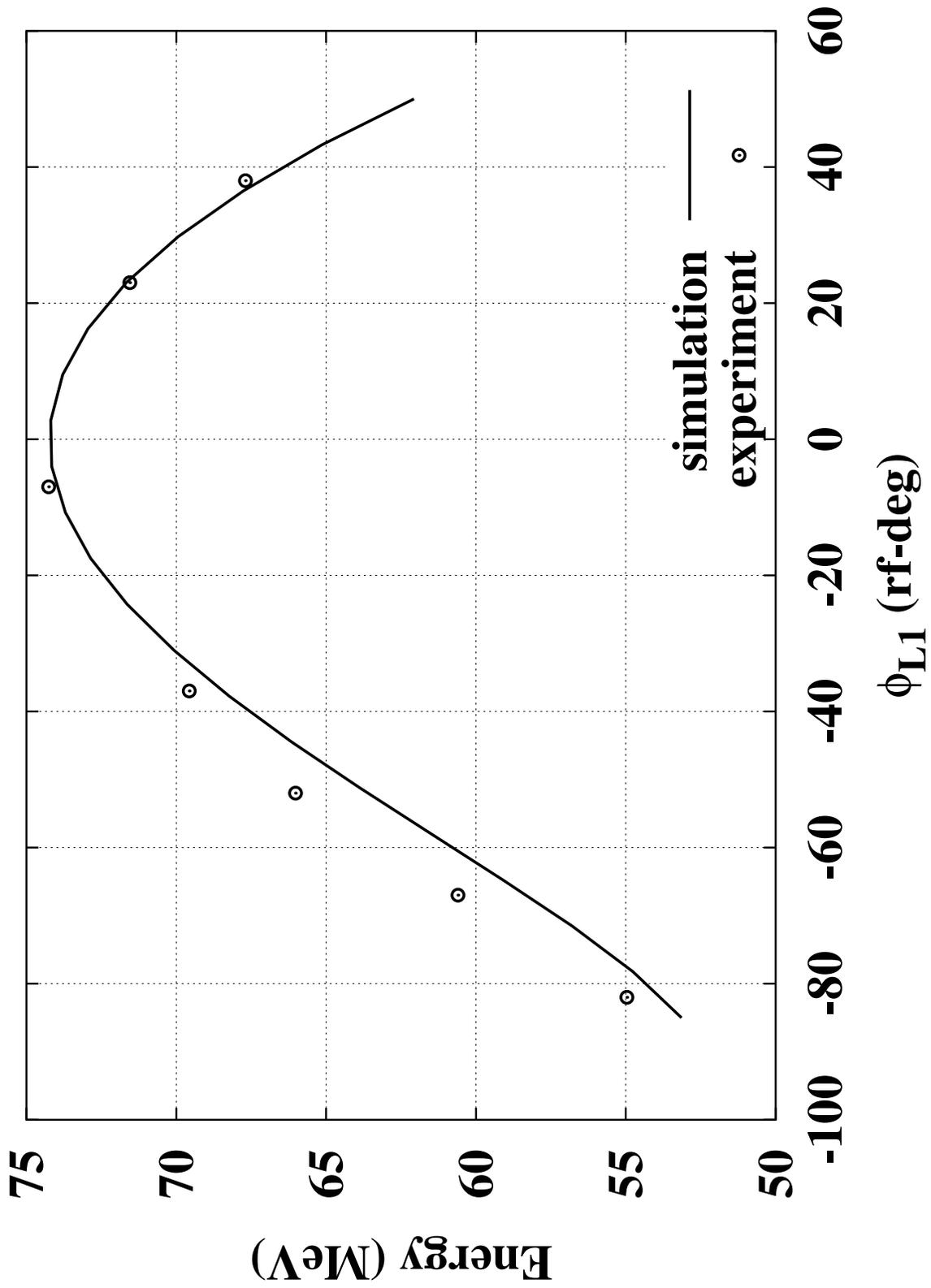}
\end{center}
\caption{ Total energy versus phase of linac section L1. The points are 
measurements obtained parasitically to the bunch length measurement. 
The solid line is a simulation result.}
\label{fig:E_vs_phiL1}
\end{figure}

\begin{figure}[h]
\begin{center}
\epsfxsize=150mm
\epsfbox{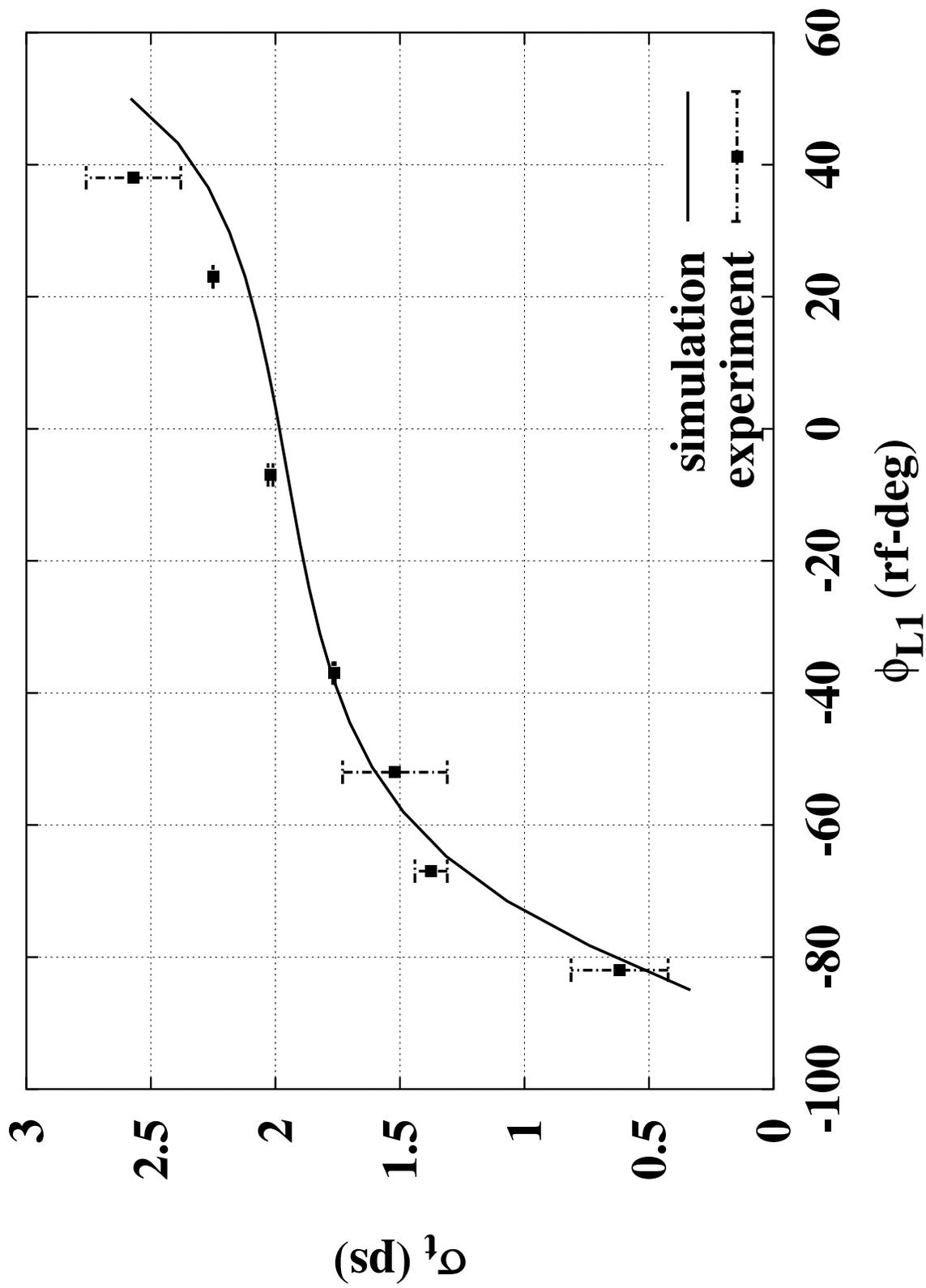}
\end{center}
\caption{ Rms bunch length versus phase of the linac section L1.}
\label{fig:bl_vs_phiL1}
\end{figure}

\begin{figure}[h]
\begin{center}
\epsfxsize=140mm
\epsfbox{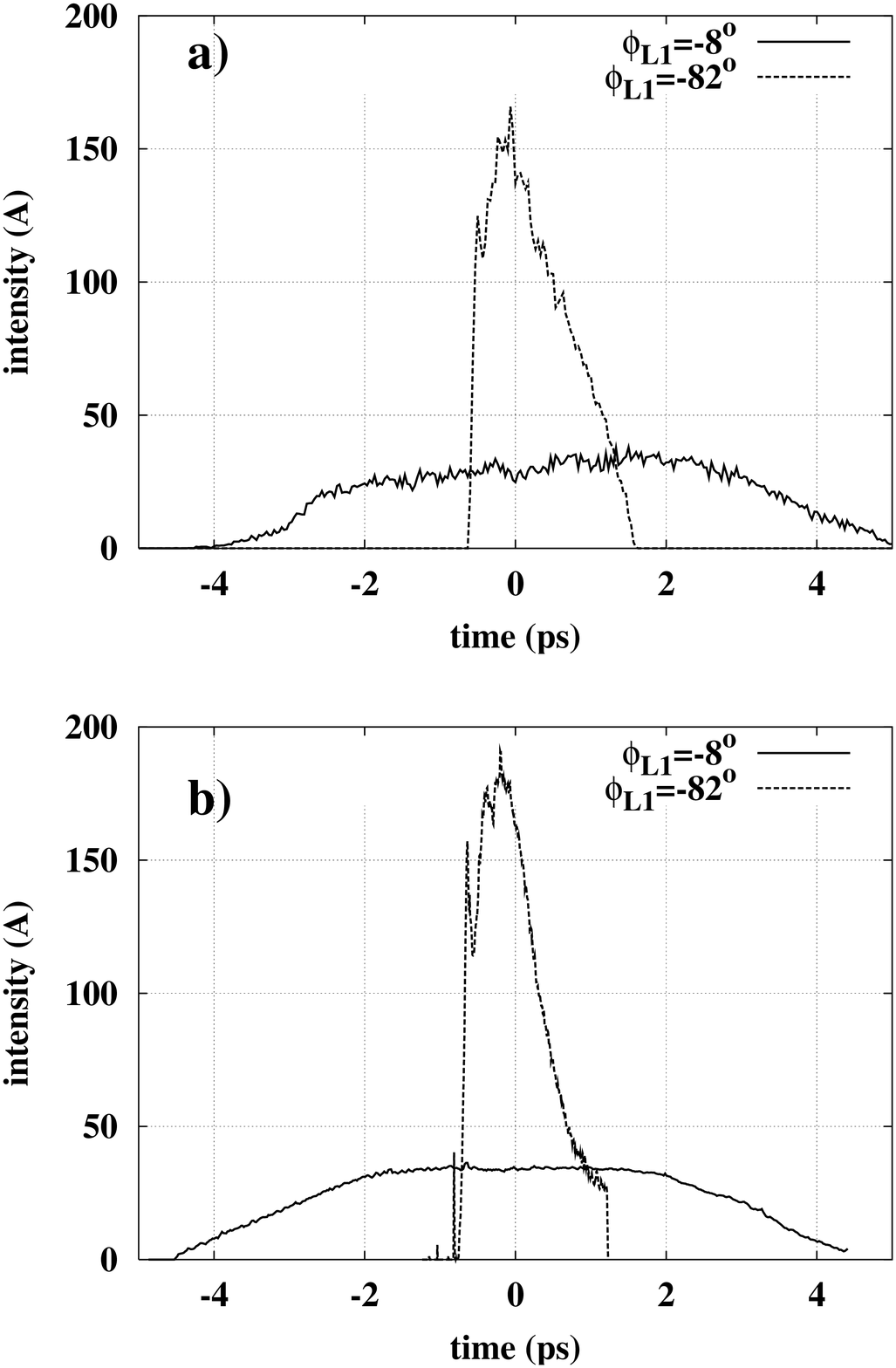}
\end{center}
\caption{ Comparison of the bunch time-profile for L1 on crest
($\phi_{L1}=0^{\circ}$), and $-82^{\circ}$ off-crest. Plot {\bf a)} was generated 
by tracking simulation; plot {\bf b)} is a direct measurement using the zero-phasing 
method. The time $>0$ corresponds to the bunch tail}
\label{fig:profile_sim_vs_meas}
\end{figure}

As the phase of the linac section L1 was varied and L2 tuned to maximize the energy gain, 
the beam energy was measured. The so-obtained energy variation versus the phase of the 
linac L1 is compared with the simulations for the nominal operating point (see Table~\ref{tab:setup})  
in Fig.~\ref{fig:E_vs_phiL1} and the corresponding plot for the bunch length is shown 
in Fig.~\ref{fig:bl_vs_phiL1}. As predicted, we observed that operating the linac at 
lower phases (thereby giving the bunch 
head less energy  than the tail) provides some compression. The parametric 
dependence of the rms bunch length on the phase of linac L1 is found to be in good 
agreement with the simulation predictions. Two cases of measured and simulated bunch 
time-profile are presented in Fig.~\ref{fig:profile_sim_vs_meas}. Again, the agreement 
between simulation and experiment is fairly good taking into account the uncertainties 
associated to the zero-phasing method. Noteworthy is the achieved peak current of 
$\sim$150~A. 


The frequency-domain technique is based on the measurement of the coherent 
radiation emitted by the electron bunch via some electromagnetic process. In the 
coherent regime (i.e. for frequencies $\omega \sim 2\pi/\sigma_t$ where $\sigma_t$ is the 
rms bunch duration) the radiated power scales with the squared charge and depends on the 
bunch form factor. Thus it provides indirect informations on the bunch time-profile. 
In DUV-FEL, we detect 
the far-field radiation associated to the geometric wake field caused by aperture 
variation along the accelerator (e.g. the irises of the rf-structure). The radiation shining 
out of a single-crystal quartz vacuum window, located prior to the linac section L3, was 
detected with a He-cooled bolometer. The detection system, composed of the bolometer and the 
vacuum extraction port, can transmit radiation within the frequency range 
$[\omega_l, \omega_u]\simeq[1.2,50]$~THz. 
The lower and upper frequency limits being respectively due to diffraction effects related 
to the finite size of the detector and transmission function of the vacuum extraction 
port. Given the bunch charge $Q$ and the Fourier transform of the bunch time-profile  
$\tilde S(\omega)$, the power is expected to scale as $P\sim Q^2 
\int_{\omega_l}^{\omega_u} d\omega |\tilde S(\omega)|^2 \propto Q^2/\sigma_t$ 
(see annex for details). The typical signal observed 
as the charge is varied is presented in Fig.~\ref{fig:coher_vs_charge}: the observed nonlinear 
behavior confirms that the emitted radiation is not incoherent. From simulation we 
expect the power to scale as $P \propto Q^{1.37}$ (see annex for details) a number close 
to the one resulting from the fit of the data:  $P \propto Q^{1.57}$.

\begin{figure}[h]
\begin{center}
\epsfxsize=150mm
\epsfbox{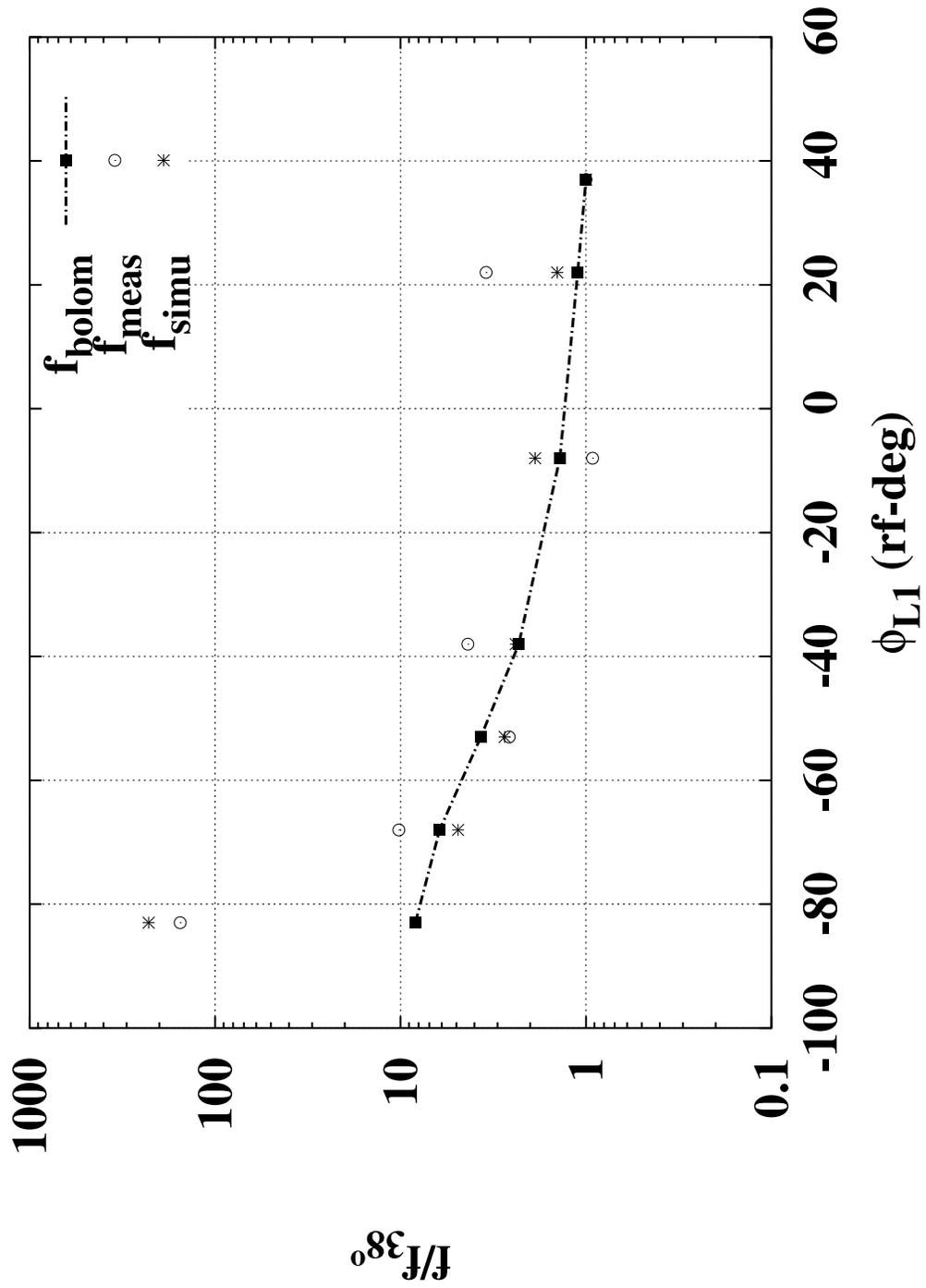}
\end{center}
\caption{ Integrated bunch form factor $f$ normalized to its 
value at $\phi_{L1}=38^{\circ}$. $f_{bolom}$, $f_{meas}$, and $f_{simu}$ 
correspond respectively to measurement with the bolometer, computation 
from the measured time-profiles and computation from the simulation-generated 
time profiles.}
\label{fig:bolom_compare}
\end{figure}

In Figure~\ref{fig:bolom_compare}, the measured bolometer output signal versus the phase 
of L1 is compared with the expectation (1) calculated from the simulated phase space density 
and (2) computed from the measured bunch time profile obtained by zero-phasing. As expected 
the increase of the coherent signal is an unambiguous signature of the bunch being compressed 
(the charge was monitored during the measurement and remained constant to 200$\pm$20~pC). \\
The data points computed from the measured time profile $f_{meas}$ were obtained by 
numerically computing the Fourier transform of the bunch time-profile (using a FFT 
algorithm) and by performing the integration:
\begin{eqnarray}
f_{meas}=\int_{\omega_l}^{\omega_u} d\omega |\tilde S(\omega) \times R(\omega)|^2 \mbox{.}
\end{eqnarray}
where $R(\omega)$ stands for the frequency response of the detection system.\\
\\
To generate the data points from the simulated phase space distributions $f_{simu}$ we 
write the time-profile, $S(t)$ as a Klimontovitch distribution:
\begin{eqnarray}\label{eqn:klimonto}
S(t)=\frac{1}{N} \sum_{i=1}^{N} \delta(t_i-t) \mbox{,}
\end{eqnarray}
$N$ being the number of macro-particle used (50000 in the simulations presented 
hereafter) and $t_i$ the time of arrival of the $i$-th macro-particle. 
Eq. (\ref{eqn:klimonto}) allows to write the integrated power as:
\begin{eqnarray}
f_{simu}=\frac{1}{N^2} \int_{\omega_l}^{\omega_u} d\omega |R(\omega)|^2 \left( \left[\sum_{i=1}^{N} 
\cos(\omega t_i) \right]^2
+ \left[\sum_{i=1}^{N} \sin(\omega t_i) \right]^2 \right)\mbox{.}
\end{eqnarray}
Though Figure~\ref{fig:bolom_compare} shows the signal increases as the bunch is 
compressed, there are discrepancies between the measurement and the two 
calculations for the short bunch case, we believe this is due to the lack of a precise 
knowledge of the transmission line frequency response. 
%
%

\section{Conclusion}
We have measured the bunch length dependence on the phase of a traveling wave accelerating
structure located just downstream from an rf electron source. We could compress the bunch by a 
factor >3, down to $\sim$0.5~ps, for a bunch charge of 200~pC. In our experimental setup, a 
stronger compression is currently difficult to achieve without significantly impinging the 
transverse phases-pace quality. The linac section used for the compression also plays a crucial 
role in achieving low emittance since it quickly accelerates the beam at energies of 
approximately  60~MeV thereby freezing the transverse phase space. Hence operating the 
first linac far off-crest reduces the final energy and impact the emittance since 
transverse space charge forces scale as $1/\gamma^2$. An improvement of our experiment would 
be to surround the accelerating structure used as a bunch compressor with a solenoidal lens 
to enable a better control of the beam transverse envelope and emittance~\cite{bacci,boscolo}.

\section{Acknowledgments}
This work was sponsored by US-DOE grant number DE-AC02-76CH00016 and by 
the Deutsches Elektronen-Synchrotron institute. We are indebted to Luca 
Serafini of Univ. Milano for carefully reading and commenting the manuscript.

\newpage
\section*{Appendix: Dependence of radiated power on bunch charge}
\begin{figure}[h]
\begin{center}
\epsfxsize=150mm
\epsfbox{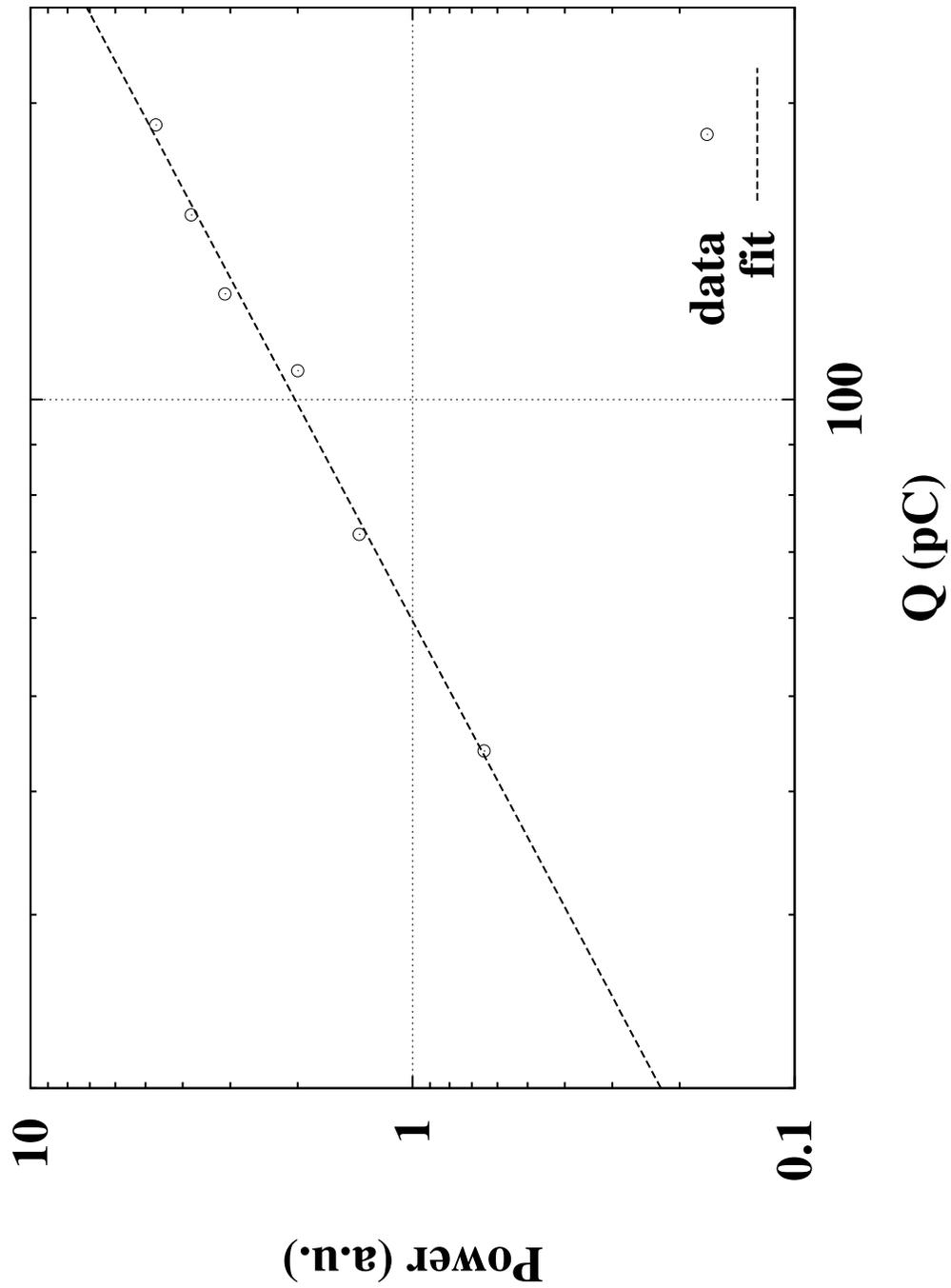}
\end{center}
\caption{ Dependence of bolometer signal versus bunch charge. The circles are 
measurement, the line is a fit of the measurement using a $\alpha \times Q^{\beta}$ 
law, the result gives $\beta=1.37\pm0.06$.}
\label{fig:coher_vs_charge}
\end{figure}

\begin{figure}[h]
\begin{center}
\epsfxsize=150mm
\epsfbox{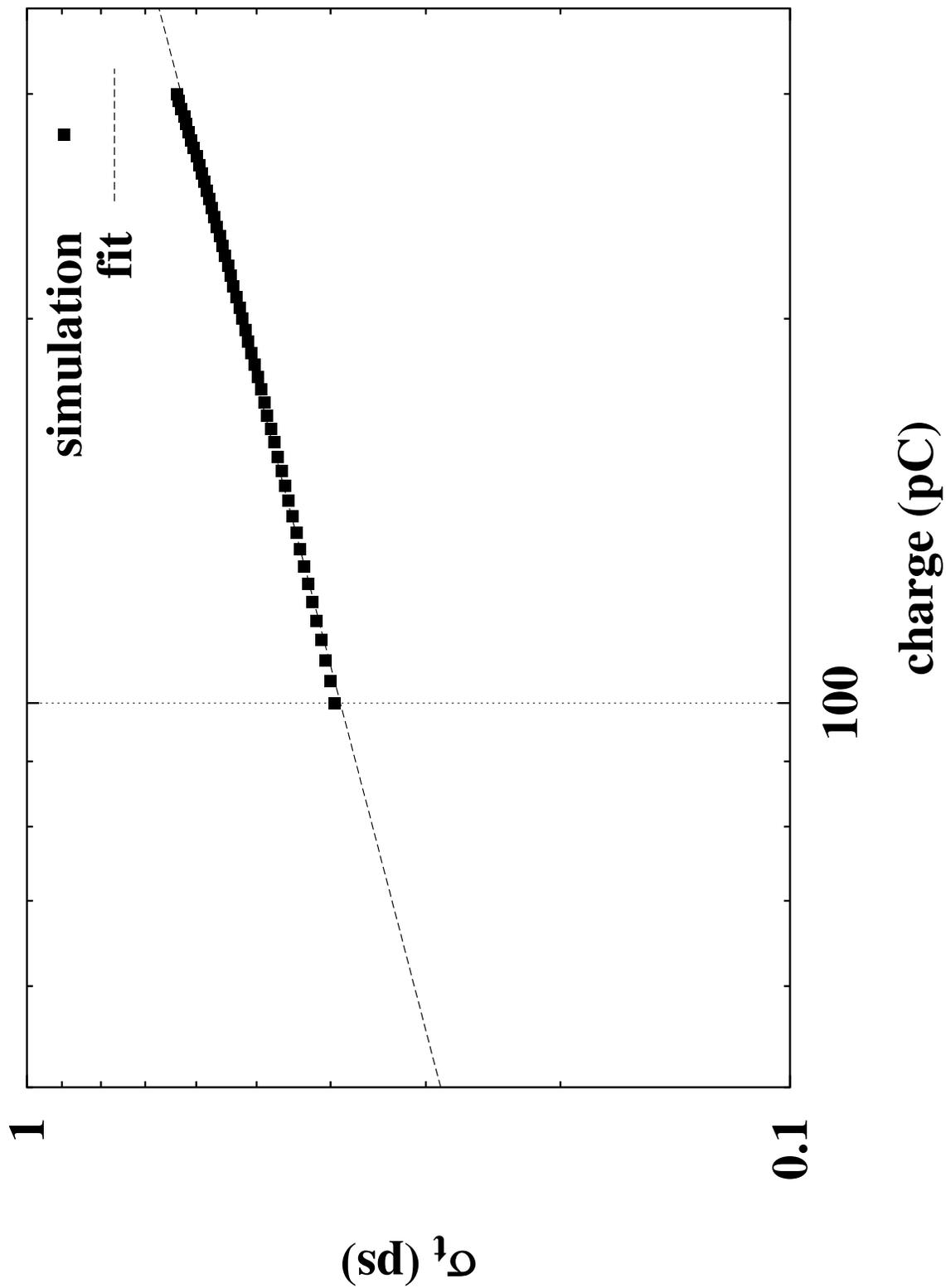}
\end{center}
\caption{ Simulated dependence of bunch length versus charge. The circles are 
simulation results, the line is a fit to the simulation using a $\alpha \times Q^{\beta}$ 
law, the result gives $\beta=0.437\pm0.007$.}
\label{fig:sigmt_vs_Q}
\end{figure}

Let's consider the case of a Gaussian distribution: 
\begin{eqnarray}
S(t)=\frac{1}{\sqrt{2\pi \sigma_t^2}} \exp \left( \frac{-t^2}{2\sigma_t^2} \right) .\nonumber
\end{eqnarray}
The corresponding bunch form factor takes the form:
\begin{eqnarray}
|S(\omega)|^2=|\int_{-\infty}^{+\infty} S(t) \exp  {-i \omega t}|^2 = \exp \left(-\sigma^2_t
\omega^2 \right),\nonumber
\end{eqnarray}
and the integrated bunch form factor in the $[\omega_l, \omega_u]$ frequency interval is:
\begin{eqnarray}
f =\int_{\omega_l}^{\omega_u} d\omega \exp \left( -\sigma_t^2 \omega^2 \right). \nonumber
\end{eqnarray}
The integration of the latter equation can be written in term of ``error" function:
\begin{eqnarray}
f =\frac{\sqrt{\pi}}{2 \sigma_t} \left( \mbox{erf} (\sigma_t \omega_u) -
\mbox{erf} (\sigma_t \omega_l) \right).
\end{eqnarray}
Taking into account the limit of the erf function,
\begin{eqnarray}
\lim_{z\rightarrow \infty} \mbox{erf}(z) = 1 \mbox{, and}\\ \nonumber
\lim_{z\rightarrow 0} \mbox{erf}(z) = 0 \mbox{,}\\ \nonumber
\end{eqnarray}
and assuming the frequency range is so that $\sigma_t \omega_u \gg 1$ and 
$\sigma_t \omega_l \ll 1$, we finally have for the radiated power:
\begin{eqnarray}
P\propto Q^2 \times f \propto \frac{Q^2}{\sigma_t} \mbox{.}
\end{eqnarray}
Figure~\ref{fig:sigmt_vs_Q} shows the dependence of the bunch length versus the charge 
expected from simulations . We find $\sigma_t \propto Q^{0.43}$ and thus we would expect 
the radiated power to be $P\propto Q^{1.57}$ which is close to the value deduced from the 
fit of the measurement presented in Fig.~\ref{fig:coher_vs_charge}: $P\propto Q^{1.37}$. 


%
%

\newpage

\newpage

\newpage

\newpage

\newpage

\newpage

\newpage
\newpage

\newpage

\end{document}